\documentclass[10pt, pre, twocolumn, showpacs, aps]{revtex4-1}
\usepackage{amsmath,amssymb,subfigure}
\usepackage[dvipdfmx]{graphicx,color}
\usepackage{verbatim}

\newcommand{\R}{\mathbb{R}}
\newcommand{\dfracp}[2]{\dfrac{\partial #1}{\partial #2}}
\newcommand{\dfracd}[2]{\dfrac{{\rm d} #1}{{\rm d} #2}}

\newcommand{\Fasym}{F_{\alpha}^{\rm asym}}
\newcommand{\Hasym}{H_{\alpha}^{\rm asym}}

\newcommand{\ave}[1]{\langle #1 \rangle}
\newcommand{\wt}{\widetilde}

\newcommand{\Kac}{K_{\alpha}^{\rm c}}

\begin{document}
\title{Discontinuous codimension-two bifurcation in a Vlasov equation}

\author{Yoshiyuki Y. Yamaguchi}
\affiliation{Department of Applied Mathematics and Physics,
  Graduate School of Informatics, Kyoto University, Kyoto 606-8501, Japan}

\author{Julien Barr{\'e}}
\affiliation{Institut Denis Poisson, Universit{\'e} d'Orl{\'e}ans, CNRS, Universit{\'e} de Tours, France}

\begin{abstract}
  In a Vlasov equation, the destabilization of a homogeneous stationary state is typically described by a continuous bifurcation characterized by strong resonances between the unstable mode and the continuous spectrum. However, when the reference stationary state has a flat top, it is known that resonances drastically weaken, and the bifurcation becomes discontinuous. In this article, we use a combination of analytical tools and precise numerical simulations to demonstrate that this behavior is related to a 
  codimension-two bifurcation, which we study in details.
\end{abstract}
\maketitle

\section{Introduction}
\label{sec:introduction}

Vlasov and other similar equations are infinite dimensional Hamiltonian systems (see for instance \cite{Morrison}) which are fundamental in many domains 
governed by long-range interactions:
plasma physics, astrophysics, fluid dynamics for instance. Getting a qualitative understanding of Vlasov dynamics is thus an old problem, which started with Vlasov and Landau \cite{Vlasov,Landau}. We shall approach this question with dynamical systems tools, in particular bifurcation theory: the rationale is that bifurcations have a universal character, and tend to provide informations on the structure of the phase space, in a sometimes rather wide neighborhood of the critical point.  

The bifurcation theory of Vlasov and Vlasov-like equations is very different from that of dissipative nonlinear partial differential equations (PDEs).
The paradigmatic case for a bifurcation of Vlasov equation is a homogeneous stationary solution with a certain velocity profile $F(p)$ which becomes unstable as a parameter (a coupling constant for instance) is varied. This situation is now well understood: the unstable eigenvalue appears embedded in the marginally stable (purely imaginary) continuous spectrum, and a reduced description involving a finite dimensional central manifold is not possible. Instead, the development and saturation of the instability is generically described by the Single Wave Model, which is itself a nonlinear PDE \cite{ONeil-Winfrey-Malmberg-71,delCastilloNegrete-98,Balmforth-Morrison-Thiffeault-13,ElskensBook}. In particular, the bifurcation is continuous, and if $\lambda$ is a real eigenvalue and indicates the instability rate, the nonlinear saturation amplitude of the instability is the peculiar $O(\lambda^{2})$ "trapping scaling", rather than the much larger $O(\lambda^{1/2})$ typical for standard pitchfork bifurcations \cite{Crawford-94,Crawford-95} in dissipative systems.      

Beyond this generic scenario, it is also well known that modifying the velocity profile of the stationary state may have a strong influence on the type of bifurcation: indeed, for "flat-top" velocity profiles, or waterbags, resonance effects between the unstable mode and the continuous spectrum are suppressed, and the validity of the standard central manifold approach is recovered; a finite dimensional reduction is then achievable, and, in all cases in which the computation has been attempted, it predicts a discontinuous bifurcation \cite{Balmforth-12,Balmforth-Morrison-Thiffeault-13}.

At the critical point, a purely imaginary eigenvalue $\lambda_I$ appears; this requires that the first derivative of the velocity profile vanishes at $\lambda_I$: $F'(\lambda_I)=0$. The generic scenario then corresponds to $F''(\lambda_I)\neq 0$, and the "flat-top" case to the vanishing of all derivatives: $F^{(n)}(\lambda_I)=0$  for any $n \in \mathbb{N}$. In the review \cite{Balmforth-Morrison-Thiffeault-13}, section VIII-C, the authors numerically analyze, in the simple setting of the Heisenberg Mean Field (HMF) model, how is the standard Single Wave Model bifurcation modified when the critical velocity profile interpolates between a gaussian and a waterbag. We undertake in this article a systematic study of this situation and show it can be understood as the influence of a special point in the family of Single Wave Model bifurcations, i.e. a kind of codimension-two bifurcation, which rules the dynamics in its neighborhood.

A typical example of codimention-two bifurcation is 
the Bogdanov-Takens bifurcation in a dissipative ordinary differential equation
\cite{Wiggins}.
Another physically important example is a tricritical point in thermodynamics; such a tricritical point has also been observed in a Vlasov system \cite{Antoniazzi-etal-07} in relation with Lynden-Bell statistical mechanics.
At variance with \cite{Antoniazzi-etal-07}, which uses non stationary waterbags initial states, we consider
in the present work small perturbations of smooth stationary reference states. 
Beyond the case homogeneous states, bifurcations of Vlasov equations have also been studied for families of nonhomogeneous (position depending) distributions,
in the context of self gravitating systems \cite{Palmer}, and  more recently in 
\cite{Barre-Metivier-Yamaguchi-16,Barre-Metivier-Yamaguchi-20}; these studies are restricted however to codimension-one bifurcations.

To be more precise, we restrict for simplicity to one-dimensional Vlasov equations
with periodic boundary condition, and to even velocity profiles. 
We consider a family $F_{\alpha}$ of stationary states
parameterized by $\alpha$,
which are unimodal for $\alpha\leq 0$ and bimodal for $\alpha>0$.
A coupling constant provides one more tunable parameter, which induces instability of the reference state,
and a codimension-two bifurcation lies on the line $\alpha=0$.
The existence of a critical unimodal velocity profile requires the interaction to be attractive, which we assume in the following.
A typical example is provided by self-gravitating systems,
and another remarkable example is a system consisting of trapped ions,
whose interaction range can be experimentally controlled
from short to long \cite{Porras-Cirac-04,Kim-etal-09,Britton-etal-12,Islam-etal-13,Richerme-etal-14}.

\begin{figure}
  \centering
  \includegraphics[width=8cm]{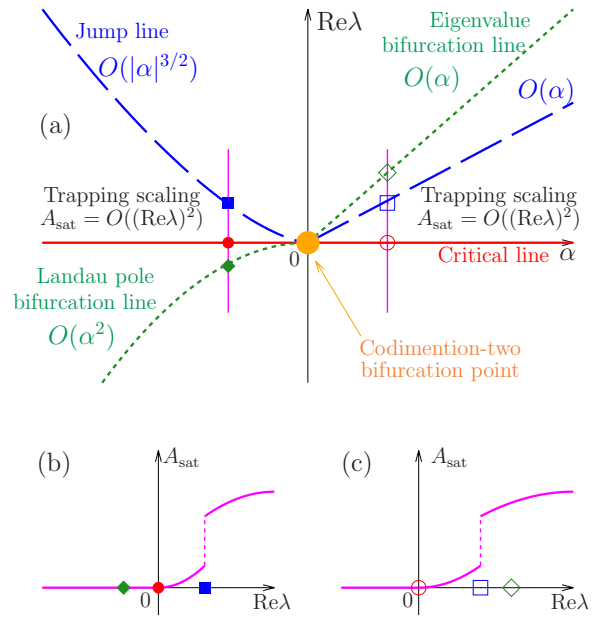}
  \caption{(a) Sketch of the two-dimensional parameter space
    $(\alpha, {\rm Re}\lambda)$,
    where $\alpha$ parameterizes a family of reference stationary states $F_{\alpha}$:
    $F_{\alpha}$ is unimodal for $\alpha\leq 0$ and bimodal for $\alpha>0$.
    $\lambda$ is the eigenvalue or Landau pole which has the largest real part.
    The codimension-two bifurcation point is the origin
    $(\alpha, {\rm Re}\lambda)=(0,0)$.
    The three types of lines are
    the critical line (red solid),
    the eigenvalue collision line (green dotted),
    and the jump line (blue dashed).
    Trapping scaling $A_{\rm sat}=O(({\rm Re}\lambda)^{2})$ appears
    between the critical line and the jump line,
    where $A_{\rm sat}$ is the asymptotically saturated amplitude of the unstable mode.
    (b) Sketch of a curve representing 
     $A_{\rm sat}$ as a function of ${\rm Re}\lambda$,
    along the left magenta vertical line on panel (a).
    (c) Same as (b) but along the right magenta vertical line.
    In both cases (b) and (c), the bifurcation is continuous with trapping scaling,
    but the asymptotic amplitude then shows a jump.
    On panel (b) [(c)], Landau damping (instability)
    is oscillatory to the left of the green diamond point,
    and nonoscillatory to the right. }
  \label{fig:intro}
\end{figure}

Our results are schematically illustrated on Fig.\ref{fig:intro}.
We first analyze the codimension-two bifurcation at the linear level,
showing that it is characterized by a collision of two complex conjugate eigenvalues (or Landau poles) $\lambda$ and $\lambda^\ast$ on the real axis.
We call this in the following {\it eigenvalue collision}; it should not be confused with 
the points where one or two eigenvalues cross the imaginary axis: at these points the reference state becomes unstable, and we call them {\it critical points}.
For simplicity, when Landau poles (and not bona fide eigenvalues) collide on the real axis, we also call it 
an {\it eigenvalue collision}.
At the codimension-two point, which we shall also call {\it bifurcation point},
the eigenvalue collision happens exactly for $\lambda=0$,
at the same time as the critical point.

In a neighborhood of the bifurcation point, Landau poles are close to the imaginary axis, and not always real: Landau damping is then weak and may be oscillating. 
As standard central manifold expansion is in general not valid in this case, we use a combination of complementary methods
to study the bifurcation at the nonlinear level:

i) The self-consistent equation \cite{Leoncini-VanDenBerg-Fanelli-09,deBuyl-Mukamel-Ruffo-11,Ogawa-Yamaguchi-14,Ogawa-Yamaguchi-15,Tacu-Benisti-22}, which focuses on computing approximately the asymptotic stationary state after the nonlinear evolution of the instability. 
It predicts a discontinuous transition at the codimension-two bifurcation point; in the unimodal region $\alpha<0$, it predicts a continuous bifurcation, followed, deeper in the unstable region, by a discontinuous jump of the asymptotic state.
However, the self-consistent equation is not applicable for the bimodal region $\alpha>0$ close to the tricritical point $\alpha=0$.

ii) Direct numerical simulations, which confirm the analytical results where they are available, and allow to explore the regimes where they are not. 
Numerical simulations reveal in particular that the bifurcation is always continuous
except at the codimension-two bifurcation point, but that this continuous bifurcation is followed by a jump of the asymptotic state
in the bimodal side $\alpha>0$ as well as the unimodal side $\alpha<0$. 
The region where the bifurcation is continuous, and which is described by trapping scaling and the Single Wave Model, drastically shrinks
when we approach the codimension-two bifurcation point from either side,
vanishing at the bifurcation point.
We also complement our analysis by studying the case of more vanishing derivatives
of the critical profile $F_{0}$.

The rest of the paper is organized to explain Fig.~\ref{fig:intro} as follows.
We present the model and the corresponding Vlasov equation
in more details in Sec.~\ref{sec:model}.
We develop the linear theory of the bifurcation in Sec.~\ref{sec:eigenvalue-bifurcation-line}.
The linear theory in particular derives the eigenvalue bifurcation point,
which plays an essential role to understand the jump in the bimodal case ($\alpha>0$).
A nonlinear theory is developed in Sec.~\ref{sec:jump} and used to analyze in details the unimodal case ($\alpha\leq 0$),
including the jump line following the continuous bifurcation.
Direct numerical simulations
of the Vlasov equation in Sec.~\ref{sec:numerics} provide comparisons and complements for these theoretical predictions.

\section{Model}
\label{sec:model}

We consider a spatially one-dimensional system with periodic boundary condition.
The $N$-body Hamiltonian is
\begin{equation}
  \label{eq:model}
  H_{N} = \sum_{i=1}^{N} \dfrac{p_{i}^{2}}{2}
  + \dfrac{1}{2N} \sum_{i=1}^{N} \sum_{j=1}^{N} \phi(q_{i}-q_{j}),
\end{equation}
where $\phi(q)$ is a $2\pi$-periodic and even coupling function.
The coupling function is then expanded in Fourier series as
\begin{equation}
  \phi(q) = - \sum_{k=1}^{\infty} K_{k} \cos kq,
  \label{eq:phi}
\end{equation}
where the constant term ($k=0$) was omitted.
A positive coefficient $K_{k}>0$ means that the $k$th Fourier mode
generates an attractive interaction,
which may destabilize the homogeneous state.
If $K_{1}=1$ and $K_{k}=0~(k>1)$,
the $N$-body system is called the Hamiltonian mean-field (HMF) model
\cite{Inagaki-Konishi-93,Antoni-Ruffo-95},
which is a paradigmatic mean-field model.
We assume that
\begin{equation}
  K_{1} > |K_{k}| \quad (k>1)
  \label{eq:ffm}
\end{equation}
so that the instability occurs in the first Fourier mode.
We shall use $K_{1}$ as the first bifurcation parameter
corresponding to $\lambda$ on Fig.~\ref{fig:intro},
and rename it $K$ for simplicity:
the homogeneous state is stable for small $K$ and unstable for large $K$.

The mean-field like interaction in \eqref{eq:model} allows to describe dynamics
of the $N$-body system in the limit $N\to\infty$ by the Vlasov equation
\cite{Braun-Hepp-77,Dobrushin-79,Spohn-91}
\begin{equation}
  \label{eq:Vlasov}
  \dfracp{f}{t} + \dfracp{H[f]}{p} \dfracp{f}{q} - \dfracp{H[f]}{q} \dfracp{f}{p} = 0.
\end{equation}
Here, $f(q,p,t)$ is the one-particle distribution function
with the normalization condition
  \begin{equation}
    \label{eq:normalization-condition}
    \iint_{\mu} f(q,p,t) dqdp = 1,
  \end{equation}
and $H[f](q,p,t)$ is the one-particle Hamiltonian functional defined by
\begin{equation}
  H[f](q,p,t) = \dfrac{p^{2}}{2} + \iint_{\mu} \phi(q-q') f(q',p',t) dq'dp',
\end{equation}
where $\mu$ is the one-particle phase space
spanned by the position variable $q\in (-\pi,\pi]$
and the conjugate momentum variable $p\in\R$.

We recall three important facts on the Vlasov equation.
First, any homogeneous distribution, which depends on $p$ only,
is a stationary solution to the Vlasov equation \eqref{eq:Vlasov}.
Second, the Vlasov equation has an infinite number of conserved quantities, called Casimir invariants,
irrespective of the Hamiltonian. A Casimir invariant is of the form
\begin{equation}
  \mathcal{C}[f] = \iint_{\mu} c(f(q,p)) dqdp,
\end{equation}
where $c$ is an arbitrary smooth function.
Third, from the condition \eqref{eq:ffm},
the stability of a homogeneous stationary state $F(p)$ is obtained
from the spectral function for the first Fourier mode, $\Lambda_{1}(\lambda)$, where
the spectral function for the $k$th Fourier mode is
\begin{equation}
  \label{eq:spectrum-function}
  \Lambda_{k}(\lambda) =1 + K_{k}\pi \int_{\R} \dfrac{F^{(1)}(p)}{p-i\lambda/k} dp.
\end{equation}
The superscript with the parentheses represents the order of the derivative:
\begin{equation}
  F^{(l)}(p) = \dfracd{{}^{l}F}{p^{l}}(p).
\end{equation}
Roots of $\Lambda_k(\lambda)$ are eigenvalues of the linearized Vlasov equation around the reference stationary state $F$.
Clearly, if there exists an eigenvalue
whose real part is positive, then $F$ is unstable. Thanks to \eqref{eq:ffm}, the destabilization of the profile $F$ occurs through the first Fourier mode. 
Hence we shall use the magnetization $M$ to quantify the instability, where
\begin{equation}
  \label{eq:MxMy}
  M_x+iM_y = Me^{i\varphi} = \iint_{\mu} e^{iq} f(q,p)dq\,dp.
\end{equation}

The second bifurcation parameter $\alpha$ is introduced as follows.
We consider a family of homogeneous stationary states $\{F_{\alpha}(p)\}_{\alpha}$,
which are even in $p$ and such that $F_{\alpha}^{(2)}(0)$ changes sign at $\alpha=0$.
For simplicity we take $\alpha$ so that
\begin{equation}
  \label{eq:alpha-def}
  \alpha = F_{\alpha}^{(2)}(0).
\end{equation}
We assume that $F_{\alpha}(p)$ is unimodal for $\alpha\leq 0$
and bimodal for $\alpha>0$.
The unimodality at $\alpha=0$ implies that $F_{0}^{(4)}(0)<0$ in general.
Higher-order flatness, i.e. vanishing of higher order derivatives at $p=0$,
will be discussed separately.
There is a critical strength of the coupling constant $K$
at which the reference state $F_{\alpha}$ changes stability.
This critical point depends on $\alpha$,  and is denoted by $\Kac~(>0)$.
We introduce the relative distance from the critical point as
\begin{equation}
  \kappa_{\alpha} = \dfrac{K-\Kac}{\Kac}.
\end{equation}

\begin{figure}[h]
  \centering
  \includegraphics[width=8cm]{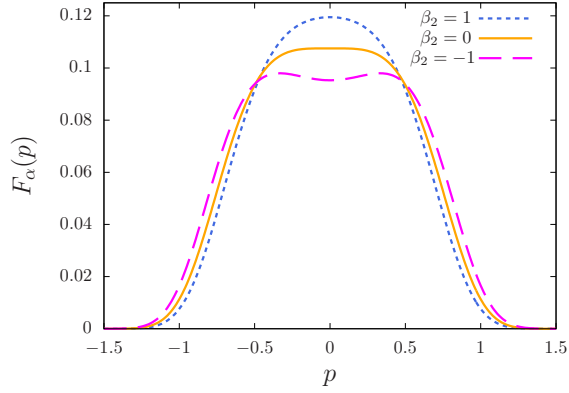}
  \caption{Examples of the reference states $F_{\alpha}(p)$ \eqref{eq:F24} with $\beta_{4}=3$.}
  \label{fig:F24}
\end{figure}

In the explicit computations of Secs.~\ref{sec:eigenvalue-bifurcation-line} and \ref{sec:numerics},
we use the family of stationary states
\begin{equation}
  \label{eq:F24}
  F_{\alpha}(p) = A \exp\left[ - \beta_{2} p^{2}/2 - \left( \beta_{4} p^{2}/2 \right)^{2} \right],
  \quad
  \beta_{4}=3,
\end{equation}
where $A$ is the normalization factor, so that $F_{\alpha}$ satisfies the normalization condition \eqref{eq:normalization-condition}.
The bifurcation parameter $\alpha$ is defined by
\begin{equation}
  \alpha = F_{\alpha}^{(2)}(0) = - A \beta_{2}.
\end{equation}
Some examples of $F_{\alpha}(p)$ are shown in Fig.~\ref{fig:F24}.

\section{Linear Theory : Eigenvalue collision}
\label{sec:eigenvalue-bifurcation-line}

The eigenvalue collision is derived from
the linear theory of the Vlasov equation.
The linearized Vlasov operator has a continuous spectrum spanning the whole imaginary axis. It may also have eigenvalues, given by the roots of the spectral functions \eqref{eq:spectrum-function}. Since the instability occurs on the first Fourier mode [thanks to condition \eqref{eq:ffm}], the $\Lambda_k$ functions for $k\neq \pm 1$ have no roots in the neighborhood of the bifurcation: indeed, the existence of an eigenvalue $\lambda$ would imply
by Hamiltonian symmetry the existence of an eigenvalue $-\lambda$,
and the reference state would be unstable.
The spectral function $\Lambda_{-1}$ is the complex conjugate of $\Lambda_{1}$,
hence we concentrate on 
\begin{equation}
  \label{eq:spectrum-function1}
  \Lambda_{1}(\lambda,\kappa_{\alpha},\alpha)
  =1 + (1+\kappa_{\alpha}) \Kac\pi \int_{\R} \dfrac{F_\alpha^{(1)}(p)}{p-i\lambda} dp.
\end{equation}

We see from this expression that $\Lambda_1$ is holomorphic on the domains ${\rm Re}\, \lambda>0$ and ${\rm Re}\, \lambda <0$, but not on the whole complex plane. 
On the stable side of the bifurcation ($\kappa_{\alpha}<0$), there are no eigenvalues;
there are however Landau poles, which are roots of the analytically continued
spectral function \eqref{eq:spectrum-function1}
from the right-half plane ${\rm Re}\lambda>0$
to the left-half plane ${\rm Re}\lambda\leq 0$.
The continuation is performed by continuously deforming the integration contour from $\R$ to a new contour L so as to avoid the singular point $p=i\lambda$,
which is in the upper-half of the complex $p$ plane for ${\rm Re}\lambda>0$, goes down on the real axis for ${\rm Re}\lambda=0$
and moves to the lower-half for ${\rm Re}\lambda<0$.
The continued integral is expressed for an analytic function $g(p)$ as
\begin{equation}
  \int_{\rm L} \dfrac{g(p)}{p-i\lambda} dp
  = \left\{
    \begin{array}{ll}
      \displaystyle{ \int_{\R} \dfrac{g(p)}{p-i\lambda} dp }
      & ({\rm Re}\lambda>0) \\
      \displaystyle{ {\rm P} \int_{\R} \dfrac{g(p)}{p-i\lambda} dp }
      + i\pi g(i\lambda) 
      & ({\rm Re}\lambda=0) \\
      \displaystyle{ \int_{\R} \dfrac{g(p)}{p-i\lambda} dp }
      +  i 2\pi g(i\lambda) 
      & ({\rm Re}\lambda<0) \\
    \end{array}
  \right.
\end{equation}
where the notation ${\rm P}\int\cdots$ stands for the Cauchy principal value.
The second term in the second and the third lines is the residue at $p=i\lambda$.

We approximately obtain an eigenvalue or a Landau pole $\lambda$
by expanding the spectral
function $\Lambda_{1}$ in a Taylor series of $\lambda$:
\begin{equation}
  \label{eq:Lambda1-expand}
  \Lambda_{1}(\lambda,\kappa_{\alpha},\alpha)
  = - (1+\kappa_{\alpha}) \left[
    a_{\alpha} + b_{\alpha} \lambda - c_{\alpha} \lambda^{2} + d_{\alpha} \lambda^{3}
    + \cdots \right],
\end{equation}
where
\begin{equation}
  \label{eq:factor-abcd}
  \begin{split}
    &
    a_{\alpha} = \dfrac{\kappa_{\alpha}}{1+\kappa_{\alpha}} - \Lambda_{1}(0,0,\alpha),
    \quad
    b_{\alpha} = \Kac \pi^{2} \alpha, \\
    & c_{\alpha} = - \dfrac{1}{2} \Kac \pi \int_{\R} \dfrac{F_{\alpha}^{(3)}(p)}{p} dp,
    \quad
    d_{\alpha} = - \dfrac{1}{3!} \Kac \pi^{2} F_{\alpha}^{(4)}(0).
  \end{split}
\end{equation}
Details of the above expansion are reported in
Appendix \ref{sec:expansion-spectrum-function}.
We assume that $c_{\alpha}>0$:
This assumption implies
\begin{equation}
  \label{eq:Lambda-00alpha}
  \left\{
    \begin{array}{ll}
      \Lambda_{1}(0,0,\alpha) = 0 & (\alpha\leq 0) \\
      \Lambda_{1}(0,0,\alpha) > 0 & (0<\alpha<\alpha_{1}) \\
    \end{array}
  \right.
\end{equation}
where $\alpha_{1}>0$ is a certain small value
(see Appendix \ref{sec:Lambda-00alpha}).
Since $\kappa_\alpha=0$ corresponds to the critical line, we see from the first equation of \eqref{eq:Lambda-00alpha} that for $\alpha\leq 0$ the critical eigenvalue crosses the imaginary axis at $\lambda=0$, and the instability is non oscillatory; from the second equation of \eqref{eq:Lambda-00alpha}, we see that for $\alpha> 0$ the critical eigenvalues cross the imaginary axis away from $\lambda=0$, and the instability is oscillatory.
The assumption $c_{\alpha}>0$ is indeed true
for the family \eqref{eq:F24} around $\alpha=0$
(see Appendix \ref{sec:c-alpha}).

It is worth commenting that, from \eqref{eq:Lambda1-expand}, \eqref{eq:Lambda-00alpha}, and the coefficient $a_{\alpha}$, we have the relation
\begin{equation}
  \label{eq:Lambda1-kappa}
  \Lambda_{1}(0,\kappa_{\alpha},\alpha) = - \kappa_{\alpha}
  \quad
  (\alpha\leq 0).
\end{equation}
For $\alpha>0$, it is reasonable to assume:
\begin{equation}
  \label{eq:Lambda1-00alpha-Oalpha}
  \Lambda_{1}(0,0,\alpha)=O(\alpha)
  \quad
  (\alpha>0).
\end{equation}
We may also assume $d_{\alpha}>0$ for sufficiently small $\alpha>0$,
since, from the unimodality hypothesis, $F_{\alpha}^{(4)}(0)<0$
when $\alpha=0$, and this inequality can be continued to small $|\alpha|>0$.

Eigenvalues (or Landau poles) satisfy the equation:
\begin{equation}
  \label{eq:eigenvalue-problem}
  a_{\alpha} + b_{\alpha} \lambda - c_{\alpha} \lambda^{2} + d_{\alpha} \lambda^{3}
    + \cdots =  0.
\end{equation}
We will use a truncated version of \eqref{eq:eigenvalue-problem}
to describe a sketch of the eigenvalue bifurcation diagram
by computing eigenvalues or Landau poles
at the eigenvalue collision point $\kappa_{\alpha}^{\rm col}$
and the critical point $\kappa_{\alpha}^{\rm c}=0$;
the order of truncation we use depends on the purpose.

The eigenvalue collision corresponds to the existence of a double root of $\Lambda_{1}$,
and it can be captured by the quadratic equation
\begin{equation}
  \label{eq:eigenvalue-problem-quadratic}
  a_{\alpha} + b_{\alpha} \lambda - c_{\alpha} \lambda^{2} = 0.
\end{equation}
The degenerate real eigenvalue $\lambda_{\alpha}^{\rm col}$ is computed as
\begin{equation}
  \label{eq:lambda-alpha-b}
  \lambda_{\alpha}^{\rm col}
  = \dfrac{b_{\alpha}}{2c_{\alpha}}
  \left\{
    \begin{array}{ll}
      <0 & (\alpha<0) \\
      = 0 & (\alpha=0) \\
      >0 & (\alpha>0) \\ 
    \end{array}
  \right.
\end{equation}
which is of order $O(\alpha)$ due to $b_{\alpha}=O(\alpha)$.
Substituting $\lambda_{\alpha}^{\rm col}$ into \eqref{eq:eigenvalue-problem-quadratic},
we have
\begin{equation}
  \dfrac{\kappa_{\alpha}^{\rm col}}{1+\kappa_{\alpha}^{\rm col}}
  = \Lambda_{1}(0,0,\alpha) - \dfrac{b_{\alpha}^{2}}{4c_{\alpha}}.
\end{equation}
Recalling \eqref{eq:Lambda-00alpha} and
the assumption $\Lambda_{1}(0,0,\alpha)=O(\alpha)$ for $\alpha>0$,
we have the following signs and scalings for the eigenvalue collision point $\kappa_{\alpha}^{\rm col}$:
\begin{equation}
  \label{eq:kappab-scalings}
  \left\{
    \begin{array}{ll}
      \kappa_{\alpha}^{\rm col}<0 \text{ and } \kappa_{\alpha}^{\rm col} = O(\alpha^{2}) & (\alpha<0), \\
      \kappa_{\alpha}^{\rm col}=0 & (\alpha=0), \\
      \kappa_{\alpha}^{\rm col}>0 \text{ and } \kappa_{\alpha}^{\rm col} = O(\alpha) & (\alpha>0). \\
    \end{array}
  \right.
\end{equation}

In order to estimate the purely imaginary critical eigenvalue $\lambda_{\alpha}^{\rm c}\in i\R$, which is embedded in the continuous spectrum,
we truncate \eqref{eq:eigenvalue-problem} at cubic order.
Substituting $\lambda_{\alpha}^{\rm c}=iy~(y\in\R)$ into
\begin{equation}
  \label{eq:eigenvalue-problem-cubic}
  a_{\alpha} + b_{\alpha} \lambda - c_{\alpha} \lambda^{2} + d_{\alpha} \lambda^{3} = 0,
\end{equation}
the imaginary part of \eqref{eq:eigenvalue-problem-cubic} gives
\begin{equation}
  \label{eq:lambda-alpha-c}
    \lambda_{\alpha}^{\rm c} = \left\{
    \begin{array}{ll}
      0 & (\alpha\leq 0), \\
      \pm i \sqrt{\dfrac{b_{\alpha}}{d_{\alpha}}} & (\alpha>0). \\
    \end{array}
  \right.
\end{equation}

\begin{figure}[htb]
  \centering
  \includegraphics[width=7.5cm]{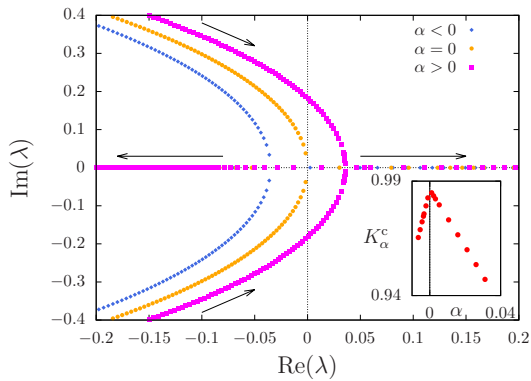}
  \caption{Collisions of eigenvalues and Landau poles
    for the family \eqref{eq:F24} close to the codimension-two bifurcation point.
    $\beta_{2}=0.3$ (unimodal $\alpha<0$, blue diamonds),
    $\beta_{2}=0$ (flat $\alpha=0$, orange circles),
    and $\beta_{2}=-0.3$ (bimodal $\alpha>0$, magenta squares)
    from left to right.
    The arrows indicate the movement of eigenvalues and Landau poles as $K$ increases.
    The inset shows the critical point $\Kac$ as a function of $\alpha$; we note an apparent singular maximum of this function at $\alpha=0$.
  }
  \label{fig:eigenvalues}
\end{figure}

For the family \eqref{eq:F24}, the eigenvalue collisions numerically computed
from the continued spectrum function are shown in Fig.~\ref{fig:eigenvalues}
with the $\alpha$ dependence of the critical point $\Kac$.
The sign of $\lambda_{\alpha}^{\rm col}$ \eqref{eq:lambda-alpha-b}
and the critical Landau pole \eqref{eq:lambda-alpha-c} are confirmed.
The scalings \eqref{eq:kappab-scalings} will be confirmed
after discussions on the trapping scaling and the jump in the nonlinearly saturated amplitude in Sec.\ref{sec:jump}.

\section{Nonlinear Theory : Trapping scaling and jump}
\label{sec:jump}

After the reference state becomes unstable,
the system reaches an asymptotic state which is close to the reference state: the bifurcation is continuous, except for $\alpha=0$. This is the region where the trapping scaling $A_{\rm sat}=O(({\rm Re}\lambda)^{2})$ is valid.
When the parameter controlling the instability is further increased, a jump in $A_{\rm sat}$ follows the continuous bifurcation.
To understand these features, we apply a nonlinear theory,
the self-consistent equation, which is a powerful tool for Vlasov and Vlasov-like equations.
We sketch the derivation of the self-consistent equation in Sec.~\ref{sec:self-consistent-equation},
and discuss the continuity of the bifurcation in Sec.~\ref{sec:continuity-bifurcation}.
For $\alpha<0$ (unimodal), we show in Sec.~\ref{sec:jump-alphanegative} that
the well-known trapping scaling $O(({\rm Re}\lambda)^{2})$
is reproduced by the self-consistent equation,
and that the scaling of the jump point $\kappa_{\alpha}^{\rm J}=O(|\alpha|^{3/2})$
is also predicted.
The self-consistent equation has a limitation:
the asymptotic state must be stationary;
this condition is not satisfied for small $\alpha>0$ (bimodal).
We therefore propose another theory to predict the scaling:
$\kappa_{\alpha}^{\rm J}=O(\alpha)$ for $\alpha>0$ in Sec.~\ref{sec:jump-alphapositive}.
The investigation of the trapping scaling for $\alpha>0$ is left for numerical examinations.

\subsection{Self-consistent equation}
\label{sec:self-consistent-equation}

The idea of the self-consistent equation is to assume that there exists an asymptotic
stationary state $\Fasym$, and make the approximation that the temporal evolution is governed
by the Hamiltonian corresponding to this asymptotic state $\Hasym=H[\Fasym]$.
Introducing the $k$th Fourier components of the density in the asymptotic state
\begin{equation}
  M_{k,x}+iM_{k,y} = \iint_{\mu} e^{ikq} \Fasym(q,p) dqdp,
\end{equation}
the asymptotic Hamiltonian is:
\begin{equation}
  \label{eq:Hayms-expansion}
  \Hasym = \dfrac{p^{2}}{2} - \sum_{k=1}^{\infty} K_{k}
  \big( M_{k,x} \cos(kq) + M_{k,y} \sin(kq) \big).
\end{equation}

The asymptotic Hamiltonian system is integrable,
so that we can introduce angle-action variables $(\theta,J)$.
The temporal dynamics driven by $\Hasym$ conserves the action and evolves linearly the angle.
The asymptotic state is then obtained by taking the average
of the initial reference state $F_{\alpha}(p)$ over the $\theta$ variable, at fixed $J$: 
\begin{equation}
  \label{eq:Fasym}
  \Fasym(J) = \dfrac{1}{2\pi} \int_{0}^{2\pi} F_{\alpha}(p(\theta,J)) d\theta
  =: \ave{F_{\alpha}}_{J},
\end{equation}
where the symbol $\ave{\cdot}_{J}$ represents the average.
The right-hand side $\ave{F_{\alpha}}_{J}$ actually depends on the asymptotic state 
through the definition of angle-action variables,
hence equation \eqref{eq:Fasym} must be solved self-consistently.
The asymptotic state \eqref{eq:Fasym} conserves all Casimir invariants
up to linear order in $\Fasym-F$, that is
\begin{equation}
  \mathcal{C}[\Fasym] - \mathcal{C}[F_{\alpha}] = O(|\Fasym-F_{\alpha}|^{2}).
\end{equation}

We start with four remarks.
First, we have to assume the existence of an asymptotic stationary state.
The bimodal case with small $\alpha>0$ is then out of scope,
since the two peaks in the velocity profile induce two resonances,
and the two resonances create two traveling clusters at opposite velocities.
This two-cluster state is not stationary.
Second, the self-consistent equation is a priori applicable for any choice of coupling function $\phi$.
However, we need to construct the angle-action variables $(\theta,J)$.
They have explicit expressions in terms of Legendre elliptic integrals for the HMF model
(see \cite{Barre-Olivetti-Yamaguchi-10} for instance),
whose one-particle dynamics is essentially a pendulum,
but we need more complicated functions for a generic $\phi$, and computations become impractical. 
Third, and notwithstanding the previous remark, one expects that the self-consistent equation captures qualitative features of a system with a generic $\phi$:
indeed the higher-order order parameters
$M_{k}=(M_{k,x}^{2}+M_{k,y}^{2})^{1/2}~(k\geq 2)$ are expected to be sufficiently small
compared to $M_{1}$ around the critical point.
Finally, although the self-consistent equation is only approximate,
it has already proved powerful to analyze the critical phenomenon,
when $|\Fasym-F_{\alpha}|$ is sufficiently small around the critical point
\cite{Yamaguchi-Ogawa-15}.

We are interested in the order parameter of the unstable mode,
$(M_{1,x},M_{1,y})$, which is denoted by $(M_{x},M_{y})$ for simplicity.
Without loss of generality, we may assume $M_{y}=0$, thanks to rotational symmetry of the system.
We also assume $M_{x}>0$ and denote $M=|M_{x}|$.
The asymptotic state $F_{\alpha}^{\rm asym}$ induces
the self-consistent equation for $M$:
\begin{equation}
  \label{eq:M1-SC}
  M = \iint_{\mu} \cos q~ \Fasym\big(J(q,p)\big) dqdp,
\end{equation}
where $\Fasym$ depends on $M$ through
the asymptotic Hamiltonian $\Hasym$.
A nonzero order parameter $M>0$ induces a separatrix on the $\mu$ space,
and the width of the separatrix is of order $O(\sqrt{M})$ in the $p$-direction.

We expand the self-consistent equation \eqref{eq:M1-SC} in a power series of $M$,
which contains half-integer powers coming from the scaling $p=O(\sqrt{M})$.
The expanded self-consistent equation is \cite{Ogawa-Yamaguchi-14}
\begin{equation}
  \label{eq:self-consistent}
  \Lambda_{1}(0,\kappa_{\alpha},\alpha) M
  = \varphi(M) M,
\end{equation}
where
\begin{equation}
  \label{eq:varphiM}
  \varphi(M)
  : = L_{3/2} M^{1/2} + L_{5/2} M^{3/2} + L_{3} M^{2} +\cdots.
\end{equation}
The coefficients $L_{3/2}$ and $L_{5/2}$ are proportional to
derivatives of $F_{\alpha}$:
\begin{equation}
  \begin{split}
    L_{3/2} & = \wt{L}_{3/2} F_{\alpha}^{(2)}(0) = \wt{L}_{3/2} \alpha, \\
    L_{5/2} & = \wt{L}_{5/2} F_{\alpha}^{(4)}(0). \\
  \end{split}
\end{equation}
In the HMF model the coefficients $\wt{L}_{3/2}$ and $\wt{L}_{5/2}$ can be computed:
\begin{equation}
  \label{eq:wtL3252}
  \wt{L}_{3/2} \simeq 5.168,
  \quad
  \wt{L}_{5/2} \simeq -0.089.
\end{equation}
The exact values above are specific of the HMF model, but 
the signs hold around the critical point for a generic system, i.e. a generic coupling function $\phi$ 
(see the third remark in Sec.~\ref{sec:self-consistent-equation}).

\subsection{Continuity of the bifurcation}
\label{sec:continuity-bifurcation}

\begin{figure}
  \centering
  \includegraphics[width=8cm]{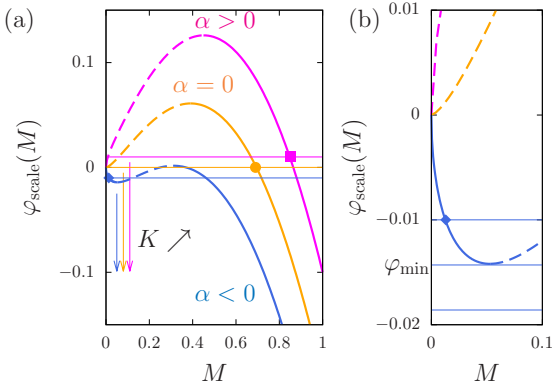}
  \caption{ (a) Schematic picture of $\varphi_{\rm scale}(M)$ \eqref{eq:varphi-scale}
    for $r=-0.1$ ($\alpha<0$, blue lower), $r=0$ ($\alpha=0$, orange middle),
    and $r=0.1$ ($\alpha>0$ magenta upper) with $\gamma=1.2$.
    A solid line represents a stable branch,
    and a dashed line an unstable branch.
    The three horizontal lines mark the level of $\Lambda_{1}(0,\kappa_{\alpha},\alpha)$,
    which goes down as the coupling constant $K$ increases
    from the critical value $K_{\alpha}^{\rm c}$.
    The three points predict the asymptotic value of $M$
    for $\alpha<0$ and $\kappa_{\alpha}>0$ (blue diamond),
    $\alpha=0$ and $\kappa_{\alpha}=0^{+}$ (orange circle),
    and $\alpha>0$ and $\kappa_{\alpha}=0^{+}$ (magenta square).
    Actually, this jump of $M$ does not happen for $\alpha>0$, see text.
    (b) Magnification of (a) around the origin.
    The middle of the three horizontal blue lines
    is the jump level at $K_{\alpha}^{\rm J}$,
    determined from $\varphi_{\rm min}$ by \eqref{eq:jump-condition-alphanegative},
    and $M$ jumps to the other stable branch of (a)
    for $K>K_{\alpha}^{\rm J}$.}
  \label{fig:SelfConsistentEq}
\end{figure}

Solutions to the self-consistent equation \eqref{eq:self-consistent}
are obtained as intersection points of the graph of $\varphi(M)$ with
the horizontal level $\Lambda_{1}(0,\kappa_{\alpha},\alpha)$,
which is a decreasing function of $\kappa_{\alpha}$ around $\alpha=0$.
To graphically understand the intersection, we consider a scaled
and truncated function $\varphi_{\rm scale}(M)$ defined by 
\begin{equation}
  \label{eq:varphi-scale}
  \varphi_{\rm scale}(M) = r M^{1/2} + M^{3/2} - \gamma M^{2},
\end{equation}
which is obtained by scaling \eqref{eq:varphiM} as
\begin{equation}
  \sqrt{M} \to \dfrac{-\gamma L_{5/2}}{L_{3}} \sqrt{M},
  \quad
  \varphi \to \dfrac{-\gamma^{3}L_{5/2}^{4}}{L_{3}^{3}} \varphi,
  \quad
  r = \dfrac{L_{3}^{2}L_{3/2}}{\gamma^{2}L_{5/2}^{3}} .
\end{equation}
Here we used the sign $L_{5/2}>0$ from $F_{0}^{(4)}(0)<0$ and continuation
around $\alpha=0$.
Moreover, we assumed that $L_{3}<0$ and $\gamma>0$
because it is the case for $F_{0}(p)$ in the HMF model
(see Appendix \ref{sec:L3-alpha}).
The sign of $r$ coincides with the sign of $\alpha$.
Graphs of $\varphi_{\rm scale}(M)$ are shown in Fig.~\ref{fig:SelfConsistentEq}
for $\gamma=1.2$.
An increasing interval of $\varphi_{\rm scale}(M)$ corresponds to an unstable branch,
because $M$ at the intersection point decreases when $\kappa_{\alpha}$ increases.

For $\alpha<0$, a stable branch exists around $M=0$
and the bifurcation is continuous.
Further increasing $\kappa_{\alpha}$, the stable branch vanishes and a jump emerges,
when the level $\Lambda_{1}(0,\kappa_{\alpha},\alpha)$ is lower than $\varphi_{\rm min}$,
which is the local minimum of $\varphi(M)$ located around $M=0$
[see Fig.~\ref{fig:SelfConsistentEq}(b)].
For $\alpha\geq 0$, there is no stable branch around $M=0$:
The self-consistent equation predicts that the bifurcation is discontinuous. 
The discontinuity for $\alpha=0$ is also predicted by the unstable
manifold expansion, reported in Appendix \ref{section:unstab_manifold}.
The discontinuity disagrees for $\alpha>0$ with Fig.~\ref{fig:intro},
and with the numerical simulations. 
There is no contradiction however: as already commented above, and as we shall see in the simulations, the asymptotic state 
for $\alpha>0$ and very close to criticality is not stationary, and is then out of scope of the self-consistent equation.

\begin{figure}
  \centering
  \includegraphics[width=8cm]{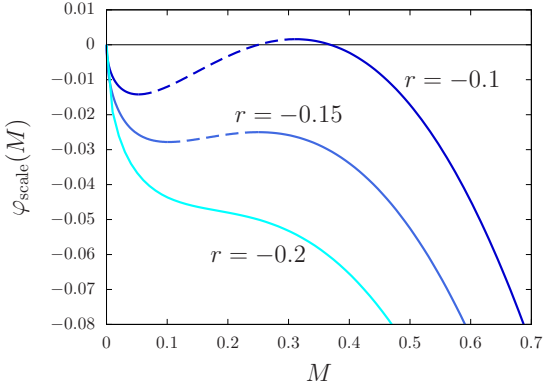}
  \caption{Graphs of $\varphi_{\rm scale}(M)$ \eqref{eq:varphi-scale} with $\gamma=1.2$.
    A solid part is a stable branch, and a dashed part is an unstable branch.
    The unstable branch and a jump disappear when $|r|$ is sufficiently large.
  }
  \label{fig:UnstableBranch}
\end{figure}

We note that smallness of $\alpha$ is crucial to have the local minimum
$\varphi_{\rm min}$ for $\alpha<0$.
Indeed, as shown in Fig.~\ref{fig:UnstableBranch},
the local minimum disappears if $|r|$ is sufficiently large.
Recalling $r=O(\alpha)$, we conclude that
the jump following a continuous bifurcation
is produced by flatness of $F_{\alpha}(p)$ around $p=0$ in $\alpha<0$ (unimodal),
and disappears for large $|\alpha|$.
This dependency on $\alpha$ is consistent with Fig.~15 of Ref.~\cite{Balmforth-Morrison-Thiffeault-13}.

We further remark that the discontinuity for $\alpha=0$ actually carries over for higher order flatness of $F_{0}(p)$:
 any $F_{0}(p)$ with a nonconstant leading term of order $O(p^{2n})~(n\geq 3)$
makes the bifurcation discontinuous, as discussed in Appendix \ref{sec:higher-order-flatness}.
An extreme case is the waterbag distribution, which is perfectly flat around $p=0$
and which is known to induce a discontinuous bifurcation \cite{Antoniazzi-etal-07}.
The above result implies that $n=2$ is sufficiently flat to make the bifurcation discontinuous.

\subsection{Trapping scaling and jump location for $\alpha<0$}
\label{sec:jump-alphanegative}

The trapping scaling $M=O\big(({\rm Re}\lambda)^{2}\big)$ is well-known,
and is reproduced by the self-consistent equation.
First, we observe the linear relation
\begin{equation}
  \label{eq:kappa-lambda}
  \kappa_{\alpha} = O({\rm Re}\lambda)
\end{equation}
from the eigenvalue problem up to the linear term:
\begin{equation}
  \label{eq:eigenvalue-problem-linear}
  a_{\alpha} + b_{\alpha} \lambda = 0,
\end{equation}
where $a_{\alpha}=\kappa_{\alpha}/(1+\kappa_{\alpha})$ for $\alpha<0$.
Second, the self-consistent equation up to the leading term of $\varphi(M)$ is
\begin{equation}
  \Lambda_{1}(0,\kappa_{\alpha},\alpha) = L_{3/2} M^{1/2}
\end{equation}
for $M>0$. The trapping scaling then results from relation \eqref{eq:Lambda1-kappa}:
\begin{equation}
  M = \left( \dfrac{\kappa_{\alpha}}{-L_{3/2}} \right)^{2} = O\big(({\rm Re}\lambda)^{2}\big).
\end{equation}

We compute now the $\alpha$ dependence of
the jump point $\kappa_{\alpha}^{\rm J}$.
The self-consistent equation has a nonzero stable solution around $M=0$
if $\Lambda_{1}(0,\kappa_{\alpha},\alpha)\geq\varphi_{\rm min}$
and loses this stable solution if $\Lambda_{1}(0,\kappa_{\alpha},\alpha)<\varphi_{\rm min}$.
The jump point $\kappa_{\alpha}^{\rm J}$ is hence computed by the equation
\begin{equation}
  \label{eq:jump-condition-alphanegative}
  \Lambda_{1}(0,\kappa_{\alpha}^{\rm J},\alpha) = \varphi_{\rm min},
\end{equation}
where, using the expansion of $\varphi$ up to order $O(M^{3/2})$:
\begin{equation}
  \varphi_{\rm min}
  = - \dfrac{2}{3} \dfrac{(-L_{3/2})^{3/2}}{(3L_{5/2})^{1/2}}.
\end{equation}
Relation \eqref{eq:Lambda1-kappa} then provides the scaling
\begin{equation}
  \label{eq:kappa-alpha-J-alphanegative}
  \kappa_{\alpha}^{\rm J}
  = \dfrac{2}{3} \dfrac{(-L_{3/2})^{3/2}}{(3L_{5/2})^{1/2}}
  = O(|\alpha|^{3/2}).
\end{equation}
The prefactor of $|\alpha|^{3/2}$ is given in Appendix \ref{sec:prefactors}.

\subsection{Scaling of the jump location for $\alpha>0$}
\label{sec:jump-alphapositive}

Since the self-consistent equation is a priori not valid in this case,
we propose a heuristic mechanism to explain the continuous bifurcation and the jump
in the bimodal case (drawing ideas from \cite{Barre-Yamaguchi-09}).
Let $\lambda$ be an eigenvalue.
The two peaks of $F_{\alpha}(p)$ create two traveling clusters
around momentum $p=\pm{\rm Im}\lambda$;
and the system may be trapped in such a non stationary bicluster asymptotic state.
The width of the clusters is of order $O(\sqrt{M})$,
which is expected to be of order $O({\rm Re}\lambda)$
from the trapping scaling $M=O\big(({\rm Re}\lambda)^{2}\big)$ (this will be checked in Sec.~\ref{sec:numerics}).
This non stationary asymptotic state is expected to disappear when the two clusters start to overlap, because this will trigger their merging;
this happens when ${\rm Im}\lambda \simeq O({\rm Re}\lambda)$.
After merging, a single cluster forms,
and the system goes to a stationary state
which is predicted by the self-consistent equation:
this is the jump.

The critical eigenvalue $\lambda_{\alpha}^{\rm c}$
and the eigenvalue at the eigenvalue collision point $\lambda_{\alpha}^{\rm col}$, corresponding by definition respectively at $\kappa_\alpha^c=0$
and $\kappa_{\alpha}^{\rm col}>0$, satisfy:
\begin{equation}
  \label{eq:scaling-lambda-cb}
  \begin{split}
    & {\rm Re}\lambda_{\alpha}^{\rm c} =0, \quad
    {\rm Im}\lambda_{\alpha}^{\rm c} = O(\sqrt{\alpha}) \\
    & {\rm Re}\lambda_{\alpha}^{\rm col} = O(\alpha), \quad
    {\rm Im}\lambda_{\alpha}^{\rm col} = 0.\\
  \end{split}
\end{equation}
We also know that ${\rm Im}\lambda$ (resp. ${\rm Re}\lambda$) is a decreasing (resp. increasing)
function of $\kappa_{\alpha}$ [see Fig.~\ref{fig:eigenvalues}(c)], and $\kappa_\alpha^{\rm col}=O(\alpha)$.

Clearly, the cluster merging condition  ${\rm Im}\lambda\simeq {\rm Re}\lambda$ is reached for $\kappa_\alpha^J$ somewhere in the interval
$0=\kappa_{\alpha}^{\rm c}<\kappa_{\alpha}^J<\kappa_{\alpha}^{\rm col}=O(\alpha)$.
Hence $\kappa_{\alpha}^J$ is at most of order $\alpha$.
Furthermore, if $\kappa_\alpha \ll \alpha$, then
\[
{\rm Re}\lambda_{\alpha} =O(\alpha)~{\rm and}~ {\rm Im}\lambda_{\alpha}=O(\sqrt{\alpha}),
\] 
so that the merging condition  ${\rm Im}\lambda \simeq {\rm Re}\lambda$ can never be met. 
We conclude that 
$\kappa_{\alpha}^J$ is of order $\alpha$, consistently with Fig.~\ref{fig:intro}.

\section{Numerics}
\label{sec:numerics}

We now illustrate and complement with detailed numerical simulations
the results of previous sections.

\subsection{The simulations setup}
We use the coupling function:
\[
\phi(q)=- \left[ K \cos(q) +K_2\cos(2q) \right], 
\]
where $K_2=0.5$ is fixed and $K$ is used as a bifurcation parameter.
We remark that $K_{2}$ is smaller than the critical point
$\Kac$ reported in the inset of Fig.~\ref{fig:eigenvalues}.
The reference family is \eqref{eq:F24}, and
\begin{equation}
  \alpha=F_{\alpha}^{(2)}(0)=-A\beta_{2}
\end{equation}
is the second bifurcation parameter. The initial condition is prepared as
\begin{equation}
  F(q,p,t=0) = F_{\alpha}(p) ( 1 + \epsilon \cos q),
\end{equation}
and the strength of perturbation is fixed as $\epsilon=10^{-6}$.

We perform numerical simulations of the Vlasov equation
by the semi-Lagrangian method described in \cite{deBuyl-10} with the timestep $\Delta t=0.05$.
The phase space $(q,p)$ is truncated as $(-\pi,\pi]\times [-4,4]$,
where the maximum value $|p|=4$ is large enough (see Fig.~\ref{fig:F24}).
We divide the phase space into an $L\times L$ mesh,
and we fix $L=512$ in the following computations.
We have checked that $L=1024$ does not significantly modify the results for $\beta_{2}=0.03$ and $0.05$.

\subsection{On the scaling relation between ${\rm Re}\lambda$ and $K-\Kac$}

The instability rate ${\rm Re}\lambda$ is commonly used as a bifurcation parameter;
for instance, the universal trapping scaling is usually expressed as
$M=O(({\rm Re}\lambda)^{2})$ on the unstable side around the critical point.
However, we will typically show curves of the magnetization as a function of the coupling constant $K$, or $\kappa_{\alpha}$.

\begin{figure}[h]
  \centering
  \includegraphics[width=8cm]{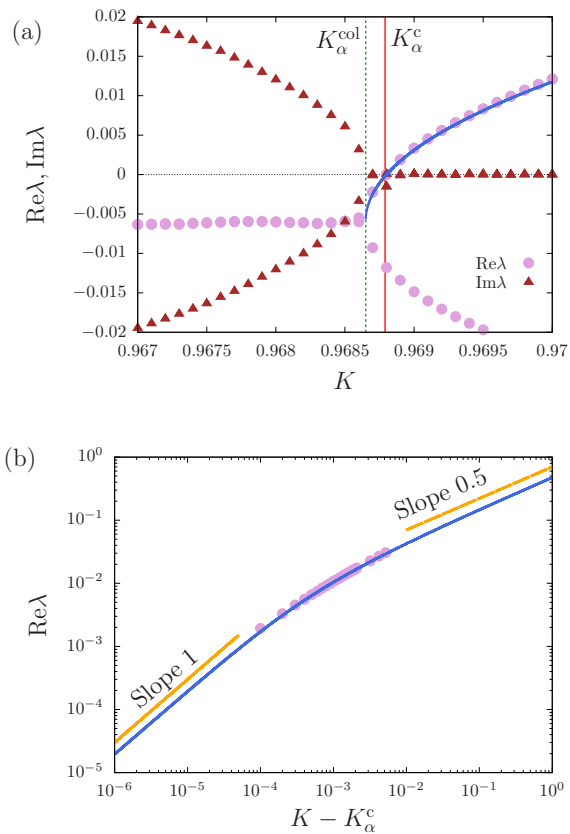}
  \caption{
    (a) Bifurcation of Landau poles. $\alpha=-0.0054$ ($\beta_{2}=0.05$).
    ${\rm Re}\lambda$ (plum circles) and ${\rm Im}\lambda$ (brown triangles)
    as functions of $K$.
    The blue solid curve represents the curve \eqref{eq:lambda-da-approximation}.
    The green dotted and red solid vertical lines mark
    the eigenvalue collision point $K_{\alpha}^{\rm col}$
    and the critical point $\Kac$ respectively.
    (b) The instability ${\rm Re}\lambda$ as a function of $K-\Kac$
    in logarithmic scale.
    The blue solid curve represents the curve \eqref{eq:lambda-da-approximation}.
  }
  \label{fig:lambda_da}
\end{figure}

In principle the choice between ${\rm Re}\lambda$ and $\kappa_{\alpha}$ is arbitrary,
as there is a linear relation between them \eqref{eq:kappa-lambda}; however, for $\alpha$ close to $0$,
this linear relation is restricted to a narrow interval of $\kappa_{\alpha}$ around $0$.
For $\alpha>0$ ($\beta_{2}<0$), the narrowness of the region is clear,
since the linear relation between ${\rm Re}\lambda$
and $\kappa_{\alpha}$ does not hold after the eigenvalue collision
$\kappa_{\alpha}>\kappa_{\alpha}^{\rm col}$,
and the eigenvalue collision point $\kappa_{\alpha}^{\rm col}$
approaches the critical point $\kappa_{\alpha}^{\rm c}=0$
as $\alpha$ goes to $0$.
For $\alpha<0$ ($\beta_{2}>0$), the narrowness of the linear region is illustrated on Fig. \ref{fig:lambda_da}.
Figure \ref{fig:lambda_da}(a) reports the bifurcation diagram of Landau poles 
for $\beta_{2}=0.05$, which corresponds to $\alpha=-0.0054$.
The unstable branch of ${\rm Re}\lambda$ is approximated by
\begin{equation}
  \label{eq:lambda-da-approximation}
  {\rm Re}\lambda
  = 0.48 \left[ \sqrt{K-K_{\alpha}^{\rm col}} - \sqrt{\Kac-K_{\alpha}^{\rm col}} \right],
\end{equation}
where
\begin{equation}
  \Kac \simeq 0.96879,
  \quad
  K_{\alpha}^{\rm col}\simeq 0.96865.
\end{equation}
Due to the smallness of $\Kac-K_{\alpha}^{\rm col}=1.5\times 10^{-4}$,
the linear region is restricted to $K-\Kac<10^{-4}$
as shown in Fig.~\ref{fig:lambda_da}(b).
Working in this region is very demanding numerically.
Therefore, we will test the trapping scaling and the jump scaling
by observing $M$ as a function of $K-\Kac$ or $\kappa_{\alpha}$
rather than of ${\rm Re}\lambda$.

\subsection{Scaling region and jump}

\begin{figure}[h]
  \centering
  \includegraphics[width=8cm]{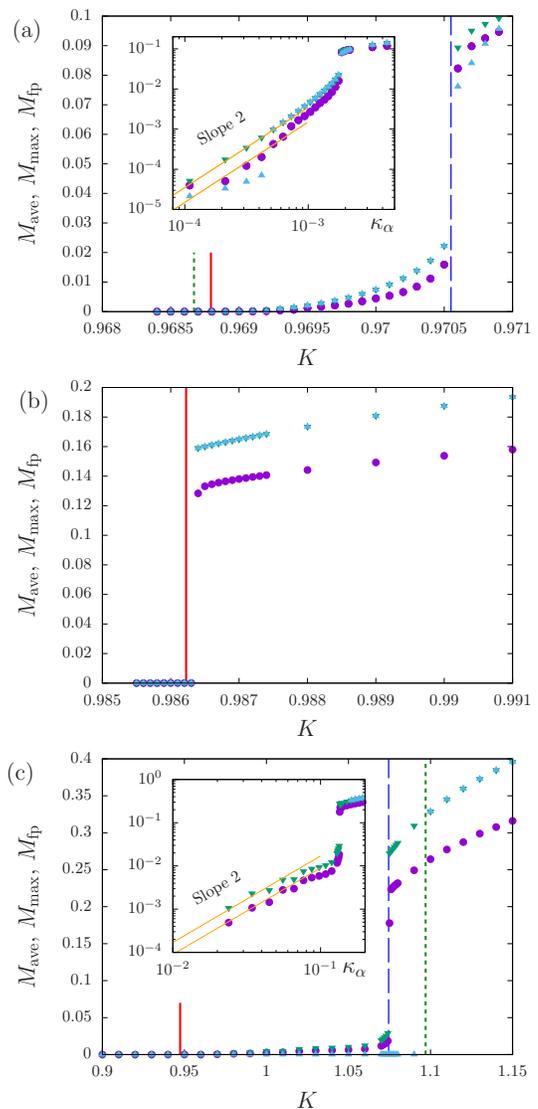}
  \caption{$M_{\rm ave}$ (purple circles),
    $M_{\rm max}$ (green inverse triangles),
    and $M_{\rm fp}$ (blue triangles)  as functions of $K$.
    (a) $\beta_{2}=0.05$ (unimodal $\alpha<0$).
    (b) $\beta_{2}=0$ (flat $\alpha=0$).
    (c) $\beta_{2}=-0.3$ (bimodal $\alpha>0$).
    In all panels, the red solid, green dotted, and blue dashed
    vertical lines represent the critical point $K_{\alpha}^{\rm c}$,
    the eigenvalue collision point $K_{\alpha}^{\rm col}$,
    and the jump point $K_{\alpha}^{\rm J}$,
    whereas the three lines coincide in the panel (b).
    In the panels (a) and (c), the insets show the three estimators
    against $\kappa_{\alpha}$ in logarithmic scale.
    The orange straight lines have slope $2$ (consistent wit trapping scaling) and are guides for the eyes.
  }
  \label{fig:M1_da}
\end{figure}

We use three estimators for the amplitude of the magnetization in the saturated state: the average
\begin{equation}
  M_{\rm ave} = \dfrac{1}{T} \int_{T/2}^{T} M(t) dt,
\end{equation}
the maximum
\begin{equation}
  M_{\rm max} = \max_{t\in [0,T]} M(t),
\end{equation}
and the first peak height $M_{\rm fp}$ of $M(t)$.
The upper limit of time is set as $T=3000$.
These estimators are shown in Fig.~\ref{fig:M1_da} as functions of $K$.
As the theory predicted, we find a jump in each panel.
The order of magnitude of the collision point $K_{\alpha}^{\rm col}$, the critical point $\Kac$,
and the jump point $K_{\alpha}^{\rm J}$ perfectly agree with Fig.~\ref{fig:intro}.
The trapping scaling $M=O(\kappa_{\alpha}^{2})$ is also confirmed
in the insets of Figs.~\ref{fig:M1_da}(a) and (c).

\begin{figure}[h]
  \centering
  \includegraphics[width=8cm]{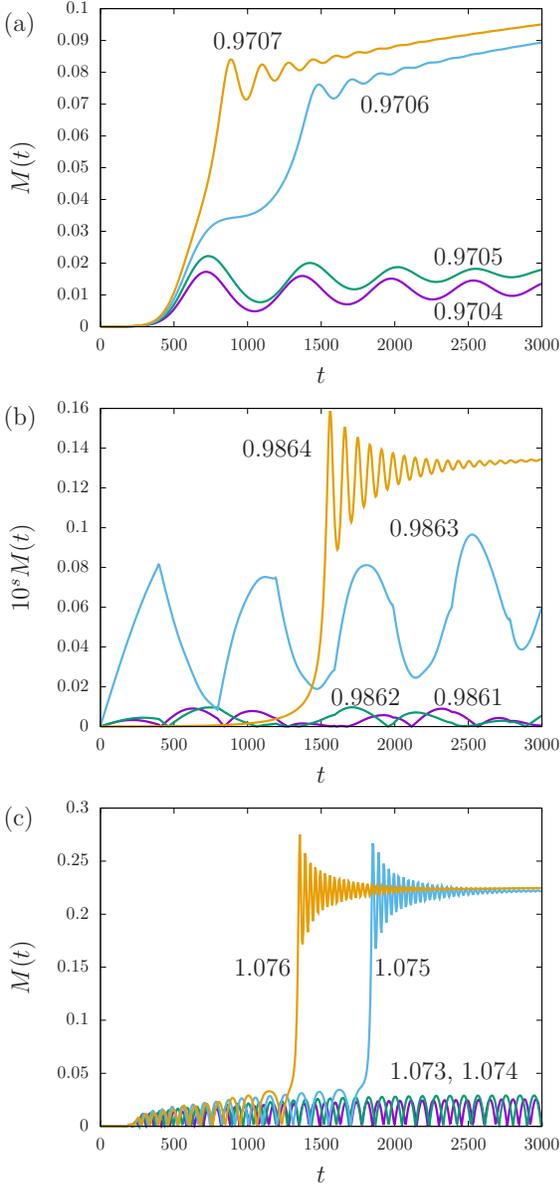}
  \caption{Temporal evolution of $M(t)$ around the jump point $K_{\alpha}^{\rm J}$.
    (a) $\beta_{2}=0.05$ (unimodal $\alpha<0$).
    (b) $\beta_{2}=0$ (flat $\alpha=0$).
    (c) $\beta_{2}=-0.3$ (bimodal $\alpha>0$).
    The numbers in the panels represent the value of $K$.
    The magnetization $M(t)$ is scaled to $10^{s}M(t)$ in the panel (b):
    $s=2$ for $K=0.9861$ and $0.9862$,
    $s=3$ for $K=0.9863$, and $s=0$ for $K=0.9864$.
  }
  \label{fig:M1t}
\end{figure}

The existence of a jump is directly confirmed from the temporal evolution of $M(t)$,
which is reported in Fig.~\ref{fig:M1t} around the jump point $K_{\alpha}^{\rm J}$.
Note that in Fig.~\ref{fig:M1t}(b) $M(t)$ is very small for $K=0.9863>K_{\alpha}^{\rm J}$, but this is caused by the slow dynamics
around the critical point. Indeed, $M(t)$ tends to slowly increase.
We remark that the slow dynamics induces a small gap between
the critical point $\Kac$ and the jump point $K_{\alpha}^{\rm J}$
in Fig.~\ref{fig:M1_da}(b).

\begin{figure}[h]
  \centering
  \includegraphics[width=8cm]{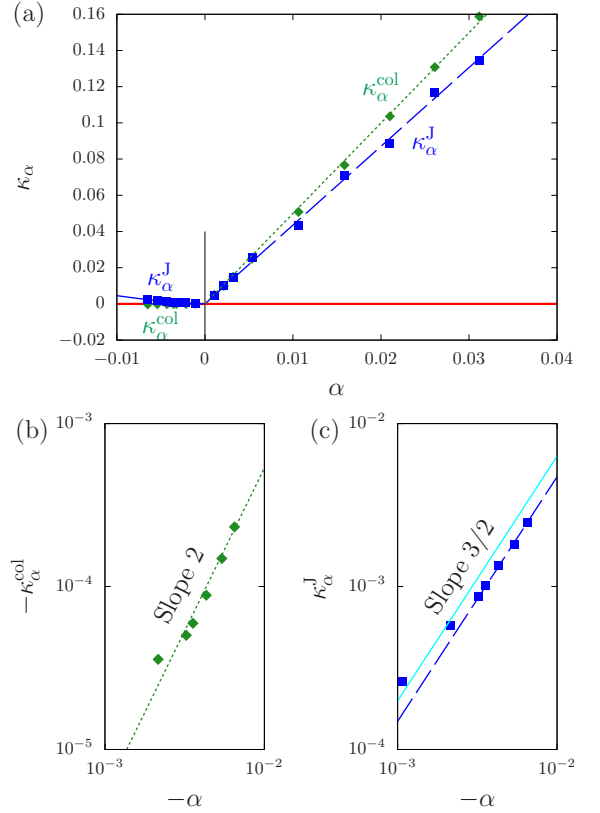}
  \caption{(a) Numerically obtained phase diagram on the plane $(\alpha, \kappa_{\alpha})$,
    which corresponds to Fig.~\ref{fig:intro}(a).
    The red solid line is the critical line.
    The eigenvalue collision point $\kappa_{\alpha}^{\rm col}$ (green diamonds)
    and the jump point $\kappa_{\alpha}^{\rm J}$ (blue squares).
    (b) Scaling of the eigenvalue collision for $\alpha<0$
    with the theoretical line $-\kappa_{\alpha}^{\rm col}=5.34\alpha^{2}$ (green dotted).
    (c) Scaling of the jump for $\alpha<0$
    with the theoretical line $\kappa_{\alpha}^{\rm J}=6.29|\alpha|^{3/2}$ (light-blue solid),
    while the estimated line has the prefactor $4.71$ (blue dashed).
  }
  \label{fig:PhaseDiagramNum}
\end{figure}

A numerically obtained phase diagram is reported in Fig.~\ref{fig:PhaseDiagramNum}(a),
which is quantitatively in good agreement with Fig.~\ref{fig:intro}(a).
For $\alpha>0$, Fig.~\ref{fig:PhaseDiagramNum}(a) verifies
the linear scaling of the eigenvalue collision $\kappa_{\alpha}^{\rm col}=O(\alpha)$
\eqref{eq:kappab-scalings} and
of the jump $\kappa_{\alpha}^{\rm J}=O(\alpha)$ (sec. \ref{sec:jump-alphapositive}).
For $\alpha<0$, Figs.~\ref{fig:PhaseDiagramNum}(b) and (c)
confirm respectively the collision scaling $\kappa_{\alpha}^{\rm col}=5.34\alpha^{2}$
\eqref{eq:kappab-scalings}
and the jump point scaling $\kappa_{\alpha}^{\rm J}=6.29|\alpha|^{3/2}$
\eqref{eq:kappa-alpha-J-alphanegative},
although the theoretical prefactor $6.29$ is somewhat larger than
the numerically obtained value $4.71$ (a similar effect is seen in \cite{Ogawa-Yamaguchi-14}).
See Appendix \ref{sec:prefactors} for the computation of theoretical prefactors for $\alpha<0$.

\subsection{Existence of two traveling clusters}

\begin{figure}[h]
  \centering
  \includegraphics[width=8cm]{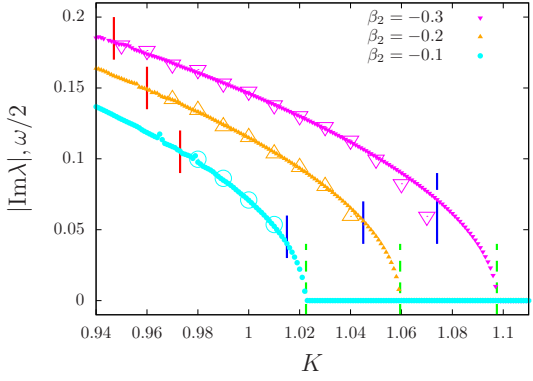}
  \caption{
    Comparison between $|{\rm Im}\lambda|$ (small symbols)
    and $\omega/2$ (large symbols),
    where $\omega$ is estimated from a time series of $M(t)$.
    $\alpha=-A\beta_{2}>0$:
    $\beta_{2}=-0.1$ (light blue circles), $-0.2$ (orange triangles),
    and $-0.3$ (magenta inverse triangles).
    Red, blue, and green vertical segments mark
    the critical point $K_{\alpha}^{\rm c}$, the jump point $K_{\alpha}^{\rm J}$,
    and the collision point $K_{\alpha}^{\rm col}$, respectively.
  }
  \label{fig:TwoClusters}
\end{figure}

Finally, we examine the existence of two traveling clusters for $\alpha>0$
in the interval between $\Kac$ and $K_{\alpha}^{\rm J}$. These clusters are very small and cannot be observed directly on the phase space density.
Instead we observe the angular frequency $\omega$ of $M(t)$,
which is extracted as the peak position of the power spectrum density.
A complex eigenvalue $\lambda$ induces an oscillation with angular frequency
${\rm Im}\lambda$, but the existence of the two traveling clusters at
$p=\pm{\rm Im}\lambda$ induces the double angular frequency
$\omega=2~{\rm Im}\lambda$.
Indeed, this relation is confirmed in Fig.~\ref{fig:TwoClusters},
which supports the existence of the two traveling clusters.

\section{Conclusions}
\label{sec:conclusions} 
We have investigated in details the bifurcation occurring in a Vlasov equation when a family of stationary states with a small curvature at the critical velocity (taken to be $0$ in this article) becomes unstable.
Our main result is
that the bifurcation of order parameter is discontinuous for the codimension-two bifurcation point where the curvature is zero,
and that away from this point and on both sides, the bifurcation is continuous and followed by a jump.
Due to this jump, the region where trapping scaling can be observed shrinks on both sides of the codimension-two bifurcation point.
Our theoretical analyses based on the self-consistent equation qualitatively predict
this phenomenology around the codimension-two bifurcation point,
and the predictions are fully confirmed by direct numerical simulations.

These results are a further step towards a classification of bifurcations in Vlasov systems \cite{Barre-Metivier-Yamaguchi-20}. Several questions remain open however. The self-consistent equation approach is restricted to the unimodal side of the bifurcation, hence our description of the bimodal side is mainly numerical. Even on the unimodal side, a better theory would be welcome; it would entail a real description of the phase space, and possibly a generalization of the Single Wave Model. This is probably challenging.

\acknowledgments

Y.Y.Y. acknowledges the support of JSPS KAKENHI
Grant Numbers 16K05472 and 21K03402.
This work has been supported by the projects RETENU ANR-20-CE40-0005-01 and PERISTOCH ANR-19-CE40-0023 of the French National Research Agency (ANR).

\appendix

\section{Linear analysis}
\label{sec:linear-analysis}

\subsection{The expansion of the spectrum function}
\label{sec:expansion-spectrum-function}

The Taylor expansion of $\Lambda_{1}(\lambda,\kappa_{\alpha},\alpha)$ is
\begin{equation}
  \Lambda_{1}(\lambda,\kappa_{\alpha},\alpha)
  = \sum_{k=0}^{\infty} \dfrac{\lambda^{k}}{k!}
  \dfracp{{}^{k}\Lambda_{1}}{\lambda^{k}}(0,\kappa_{\alpha},\alpha),
\end{equation}
where
\begin{equation}
  \dfracp{{}^{k}\Lambda_{1}}{\lambda^{k}}(\lambda,\kappa_{\alpha},\alpha)
  = i^{k} (1+\kappa_{\alpha}) \Kac \pi \int_{\R} \dfrac{F_{\alpha}^{(k+1)}(p)}{p-i\lambda} dp.
\end{equation}
Performing the analytic continuation, we have
\begin{equation}
  \label{eq:Lambda-derivatives}
  \begin{split}
    & \dfracp{\Lambda_{1}}{\lambda}(0,\kappa_{\alpha},\alpha)
    = - (1+\kappa_{\alpha}) \Kac \pi^{2} F_{\alpha}^{(2)}(0), \\
    & \dfracp{{}^{2}\Lambda_{1}}{\lambda^{2}}(0,\kappa_{\alpha},\alpha)
    = - (1+\kappa_{\alpha}) \Kac \pi \int_{\R} \dfrac{F_{\alpha}^{(3)}(p)}{p} dp, \\
    & \dfracp{{}^{3}\Lambda_{1}}{\lambda^{3}}(0,\kappa_{\alpha},\alpha)
    = (1+\kappa_{\alpha}) \Kac \pi^{2} F_{\alpha}^{(4)}(0). \\
  \end{split}
\end{equation}
The first derivative with the definition $\alpha=F_{\alpha}^{(2)}(0)$ provides
the coefficient $b_{\alpha}$,
and the second and third derivatives directly give the coefficients $c_{\alpha}$
and $d_{\alpha}$ respectively.

The constant term $a_{\alpha}$ satisfies
\begin{equation}
  \Lambda_{1}(0,\kappa_{\alpha},\alpha) = - (1+\kappa_{\alpha}) a_{\alpha}.
\end{equation}
Using the definition 
\begin{equation}
  \Lambda_{1}(0,0,\alpha) = 1 + \Kac \pi \int_{\R} \dfrac{F_{\alpha}^{(1)}(p)}{p} dp,
\end{equation}
we can modify $\Lambda_{1}(0,\kappa_{\alpha},\alpha)$ as
\begin{equation}
  \label{eq:Lambda1-kappa-alpha}
  \Lambda_{1}(0,\kappa_{\alpha},\alpha)
  = 1 + (1+\kappa_{\alpha}) [ \Lambda_{1}(0,0,\alpha) - 1 ].
\end{equation}
This modification gives the coefficient $a_{\alpha}$ of \eqref{eq:factor-abcd}.

\subsection{Spectrum function at the origin}
\label{sec:Lambda-00alpha}

We consider the spectrum function at $\lambda=0$:
\begin{equation}
  \Lambda_{1}(0,\kappa_{\alpha},\alpha)
  = 1 + (1+\kappa_{\alpha}) \Kac \pi \int_{\R} \dfrac{F_{\alpha}^{(1)}(p)}{p} dp.
\end{equation}
We show \eqref{eq:Lambda-00alpha} under the assumption $c_{\alpha}>0$.

We start from the case $\alpha\leq 0$.
At the critical point $\kappa_{\alpha}=0$,
a purely imaginary critical eigenvalue $i\lambda_{\rm I}$ (embedded into the continuous spectrum) satisfies 
\begin{equation}
  1 + \Kac \pi \left[
    {\rm P} \int_{\R} \dfrac{F_{\alpha}^{(1)}(p)}{p+\lambda_{\rm I}}
    + i\pi F_{\alpha}^{(1)}(-\lambda_{\rm I}) \right] = 0. 
\end{equation}
Considering the imaginary part of the above equation, we see that 
the unimodality of $F_{\alpha}$ implies that $\lambda_{\rm I}=0$.
Considering the real part, we then conclude $\Lambda_{1}(0,0,\alpha)=0$.

We now turn to the case $\alpha>0$.
We may assume that $|\lambda_{\rm I}|$ is small for small $\alpha>0$.
We then have the expansion
\begin{equation}
  \begin{split}
    & {\rm P} \int_{\R} \dfrac{F_{\alpha}^{(1)}(p)}{p+\lambda_{\rm I}} dp
    = {\rm P} \int_{\R} \dfrac{F_{\alpha}^{(1)}(p-\lambda_{\rm I})}{p} dp \\
    & = \int_{\R} \dfrac{F_{\alpha}^{(1)}(p)}{p} dp
    + \dfrac{\lambda_{\rm I}^{2}}{2} \int_{\R} \dfrac{F_{\alpha}^{(3)}(p)}{p} dp
    + O(|\lambda_{\rm I}|^{4}).
  \end{split}
\end{equation}
The above relation induces for $\alpha>0$ small
\begin{equation}
  \begin{split}
  \Lambda_{1}(0,0,\alpha)
  & = 1 + \Kac \pi \int_{\R} \dfrac{F_{\alpha}^{(1)}(p)}{p} dp \\
  & > 1 + \Kac \pi ~ {\rm P} \int_{\R} \dfrac{F_{\alpha}^{(1)}(p)}{p+\lambda_{\rm I}} dp = 0
  \end{split}
\end{equation}
under the assumption $c_{\alpha}>0$.

\subsection{Positiveness of the coefficient $c_{\alpha}$}
\label{sec:c-alpha}

We show that the coefficient is positive at $\alpha=0$,
namely $c_{0}>0$ for the family \eqref{eq:F24}.
Then, continuity with respect to $\alpha$ implies that
$c_{\alpha}$ is positive around $\alpha=0$.

The reference function at $\alpha=0$ is
\begin{equation}
  F_{0}(p) = A e^{-(\beta_{4} p^{2}/2)^{2}},
\end{equation}
where the normalization factor $A$ is
\begin{equation}
  A = \dfrac{1}{4\pi} \dfrac{1}{\displaystyle{\int_{0}^{\infty} e^{-(\beta_{4}p^{2}/2)^{2}} dp}}
  = \left( \dfrac{\beta_{4}}{2} \right)^{1/2} \dfrac{1}{\pi \Gamma(1/4)}
\end{equation}
and $\Gamma(z)$ is the gamma function
\begin{equation}
  \Gamma(z) = \int_{0}^{\infty} t^{z-1} e^{-t} dt.
\end{equation}
The third-order derivative of $F_{0}(p)$ is
\begin{equation}
  F_{0}^{(3)}(p)
  = - A \beta_{4}^{2} p \left( \beta_{4}^{4} p^{8} - 9 \beta_{4}^{2} p^{4} + 6 \right)
  e^{-(\beta_{4}p^{2}/2)^{2}},
\end{equation}
and the coefficient $c_{0}$ is
\begin{equation}
  \label{eq:c0}
  \begin{split}
    c_{0}
    & = \Kac \pi A \beta_{4}^{2} \int_{0}^{\infty}
    \left( \beta_{4}^{4} p^{8} - 9 \beta_{4}^{2} p^{4} + 6 \right)
    e^{-(\beta_{4}p^{2}/2)^{2}} dp \\
    & = \dfrac{\Kac \beta_{4}^{2}}{2}
    \dfrac{8 \Gamma(9/4) - 18\Gamma(5/4) + 6 \Gamma(1/4)}{\Gamma(1/4)} \\
    & = \dfrac{\Kac \beta_{4}^{2}}{2} > 0,
  \end{split}
\end{equation}
where we used the relation
\begin{equation}
  \Gamma(z+1) = z \Gamma(z).
\end{equation}

\subsection{Negativeness of the coefficient $L_{3}$}
\label{sec:L3-alpha}

We show that $L_{3}<0$ for $F_{0}(p)$ in the HMF model.
The explicit form of $L_{3}$ in the HMF model is
\begin{equation}
    \label{eq:L3}
  L_{3} = - \dfrac{5\pi}{192} \int_{\R} \dfrac{F_{\alpha}^{(5)}(p)}{p} dp,
\end{equation}
where the integral is well-defined since $F^{(5)}(p)$ is of order $O(p)$.
The fifth-order derivative of $F_{0}(p)$ is
\begin{equation}
  \begin{split}
    F_{0}^{(5)}(p)
    & = - A \beta_{4}^{4} p \left( \beta_{4}^{6} p^{14} - 30 \beta_{4}^{4} p^{10} + 195 \beta_{4}^{2} p^{6} - 210 p^{2} \right) \\
    & \times e^{-(\beta_{4}p^{2}/2)^{2}}.
  \end{split}
\end{equation}
Straightforward computations give
\begin{equation}
  \label{eq:L3-for-F0}
  L_{3} = - \dfrac{5\beta_{4}^{3}}{8} \dfrac{\Gamma(3/4)}{\Gamma(1/4)} < 0.
\end{equation}

\section{Unstable manifold expansion}
\label{section:unstab_manifold}

The idea is to set up a series expansion in powers of the amplitude of the perturbation, and to solve it order by order by projecting the full dynamics onto the unstable manifold, instead of projecting onto the central manifold as usually done; one obtains in the end a reduced equation for the amplitude, which is singular at the bifurcation point. However, it is well defined away from the bifurcation point, at variance with standard central manifold computations. By construction, it is restricted to the unstable side of the bifurcation. 
According to the study of the linearized Vlasov operator in Sec.~\ref{sec:eigenvalue-bifurcation-line}, in the unimodal $\alpha\leq 0$ case, the unstable manifold is two-dimensional, whereas it is four-dimensional in the bimodal $\alpha>0$ case. We restrict here to the unimodal case,
in which the Landau pole moves on the real axis around the critical point
[see Fig.~\ref{fig:eigenvalues}(c)].

The tangent space to the unstable manifold at the reference stationary state is spanned by the two eigenfunctions $\Phi$ and $\Phi^{\ast}$;
 we expand $f$ into
\begin{eqnarray*}
  f(q,p,t) &=& F_{\alpha}(p) + g(q,p,t), \\
\end{eqnarray*}
where
\begin{equation}
  g(q,p,t) = A(t) \Phi(q,p) + A^{\ast}(t) \Phi^{\ast}(q,p) + S(q,p,A,A^{\ast},t),
\end{equation}
and $S$ is of order $O(|A|^{2})$. The equation for the amplitude $A$ is
\begin{equation}
  \dfrac{dA}{dt} = \psi(A)
\end{equation}
where
\begin{equation}
  \psi(A) = \lambda A + c_{3}(\lambda) A |A|^{2} + O(|A|^{5})
\end{equation}
on the unstable side of the critical point, namely for $0<\lambda\ll 1$.
The coefficient $c_{3}$ is
\begin{equation}
  \label{eq:c3}
  c_{3}(\lambda) = - \left( \dfrac{\pi K}{2} \right)^{2} \wt{c}_{3}(\lambda)
\end{equation}
and
\begin{equation}
  \begin{split}
    & \wt{c}_{3}(\lambda)
    = \dfrac{1}{\lambda^{3}}
    - \dfrac{1}{\lambda^{2}} \dfrac{\Lambda_{1}^{(2)}(\lambda)}{\Lambda_{1}^{(1)}(\lambda)}
    + \dfrac{2}{3\lambda} \dfrac{\Lambda_{1}^{(3)}(\lambda)}{\Lambda_{1}^{(1)}(\lambda)}
    - \dfrac{1}{4} \dfrac{\Lambda_{1}^{(4)}(\lambda)}{\Lambda_{1}^{(1)}(\lambda)} \\
    & + \dfrac{K_{2}}{K}
    \Lambda_{1}^{(2)}(\lambda) \left[
      - \dfrac{1}{\lambda} \left( 1 + 
        \dfrac{K_{2}}{K}
        \dfrac{1}{\Lambda_{2}(2\lambda)} \right)
      + \dfrac{1}{2} \dfrac{1}{\Lambda_{2}(2\lambda)}
      \dfrac{\Lambda_{1}^{(2)}(\lambda)}{\Lambda_{1}^{(1)}(\lambda)}
    \right].
  \end{split}
  \label{eq:ctilde3}
\end{equation}
Here we omitted the arguments $\kappa_{\alpha}$ and $\alpha$ in $\Lambda_{1}$
and derivatives are performed with respect to $\lambda$.
We find a small real solution $|A|$ to the equation $\psi(A)=0$ if $c_{3}<0$,
while there is no small real solution if $c_{3}>0$.
The bifurcation is hence continuous if $\wt{c}_{3}(0)>0$,
and discontinuous if $\wt{c}_{3}(0)<0$.

The leading term of $\wt{c}_{3}$ is positive $1/\lambda^{3}$ 
when $\alpha<0$,
hence the bifurcation is continuous \cite{Crawford-94,Crawford-95}.
However, the leading singularity of $\wt{c}_{3}$ changes 
when $\alpha=0$ since $\Lambda^{(1)}(\lambda)=O(\lambda)$ from
\begin{equation}
  \Lambda_{1}^{(1)}(0,\kappa_{\alpha},0)
  = -(1+\kappa_{\alpha}) \Kac \pi^{2} \alpha = 0.
\end{equation}
With the aid of the Taylor expansions of
$\Lambda_{1}^{(1)}(\lambda,\kappa_{\alpha},\alpha)$
and $\Lambda_{1}^{(2)}(\lambda,\kappa_{\alpha},\alpha)$ around $\lambda=0$,
the leading singularity at $\alpha=0$ is
\begin{equation}
  \wt{c}_{3} \simeq \dfrac{1}{6\lambda^{2}}
  \dfrac{\Lambda_{1}^{(3)}(0)}{\Lambda_{1}^{(2)}(0)}
  = - \dfrac{1}{12\lambda^{2}}
  \dfrac{\pi F_{0}^{(4)}(0)}{\displaystyle{\int_{\R} \dfrac{F_{0}^{(1)}(p)}{p^{3}} dp }}.
  \label{eq:c3_n=2}
\end{equation}
Since the function $F_{0}^{(1)}(p)$ is of order $O(p^{3})$ around $p=0$,
the integral in the denominator is well defined.
In \eqref{eq:c3_n=2}, unimodality for $\alpha\leq 0$
implies that the numerator and the denominator are negative,
hence the bifurcation is discontinuous from $\wt{c}_{3}<0$.
We also see from \eqref{eq:ctilde3} and \eqref{eq:c3_n=2} that if $F_{\alpha}^{(2)}(0)$ is negative but small, the sign of $\wt{c}_3$ will change from positive to negative as $\lambda$ is increased from $0$ (the critical point) to some small positive value. We then expect a continuous bifurcation with trapping scaling, followed by a jump in the saturated amplitude as the distance from the instability threshold is increased: this provides a qualitative understanding to Fig.~\ref{fig:intro} (when $\alpha<0$).
We also remark that the second Fourier coefficient of the coupling function $\phi$ [see \eqref{eq:phi}]
does not affect the $\wt{c}_{3}$ factor at order $O(1/\lambda^{2})$.

\section{Discontinuity of bifurcation for higher order flatness}
\label{sec:higher-order-flatness}

At the point $\alpha=0$, the reference state is further classified
by its leading order at $p=0$.
We defined that the reference state $F(p)$ is of order $n$ when the Taylor expansion is
\begin{equation}
  F(p) - F(0) = -b p^{2n} + O(p^{2(n+1)}).
\end{equation}
We shall show now that for $F$ of order $3$ or higher, 
the self-consistent equation predicts that the bifurcation is discontinuous.

If the order of $F$ is $3$ or higher, we have $F^{(2)}(0)=F^{(4)}(0)=0$,
and hence $L_{3/2}=L_{5/2}=0$, since
\begin{equation}
  \begin{split}
    L_{3/2}
    & = F^{(2)}(0) \dfrac{M^{-3/2}}{2!} \iint_{\mu}
    \left( p^{2} \ave{\cos q}_{J} + \dfrac{M}{2} \right) dqdp \\
    L_{5/2}
    & = F^{(4)}(0) \dfrac{M^{-5/2}}{4!} \\
    & \iint_{\mu}
    \left( p^{4} \ave{\cos q}_{J} + \dfrac{M_{1}}{2} p^{2} + M^{2} \cos q \right) dqdp. \\
  \end{split}
\end{equation}
The leading term of $\varphi(M)$, \eqref{eq:varphiM},
is therefore $L_{3}$ which is
\begin{equation}
  L_{3} = - \dfrac{5\pi}{192} \int_{\R} \dfrac{F^{(5)}(p)}{p} dp.
\end{equation}
The integration is well-defined since $F^{(5)}$ is of order $O(p)$ around $p=0$.
Under the conditions $F^{(2)}(0)=F^{(4)}(0)=0$,
we can derive another expression of $L_{3}$ as
\begin{equation}
  L_{3} = - \dfrac{5\pi}{192} 4! \int_{\R} \dfrac{F^{(1)}(p)}{p^{5}} dp
\end{equation}
by repeating integration by parts,
where the integral is well-defined since $F^{(1)}$ is of order $O(p^{5})$ around $p=0$.
Therefore, we have $L_{3}>0$ for a unimodal $F$,
and the self-consistent equation $\Lambda_{1}(0)=L_{3}M^{2}$ concludes
that the bifurcation is discontinuous.
We must not confuse $L_{3}<0$ shown in Appendix \ref{sec:L3-alpha},
since the negative sign is obtained for $F^{(2)}(0)=0$ but $F^{(4)}(0)<0$,
while the positive sign is for $F^{(2)}(0)=F^{(4)}(0)=0$.
In general $L_{3}$ is not zero however high the order of $F$ is, 
hence the self-consistent equation predicts a discontinuous bifurcation for any $F$ of order $3$ or higher.

\section{Prefactors of scaling relations for $\alpha<0$}
\label{sec:prefactors}

We compute here the eigenvalue collision point $\kappa_{\alpha}^{\rm col}$
and the jump point $\kappa_{\alpha}^{\rm J}$ for the family \eqref{eq:F24}.
The theoretically obtained prefactors are used in Fig.~\ref{fig:PhaseDiagramNum}.

The eigenvalue collision point $\kappa_{\alpha}^{\rm col}$ satisfies
\begin{equation}
  \dfrac{\kappa_{\alpha}^{\rm col}}{1+\kappa_{\alpha}^{\rm col}}
  = -\dfrac{b_{\alpha}^{2}}{4c_{\alpha}}.
\end{equation}
Recalling $b_{\alpha}=\Kac\pi^{2}\alpha$, we have at leading order in $\alpha$
\begin{equation}
  \kappa_{\alpha}^{\rm col} = - \dfrac{(K_{0}^{\rm c}\pi^{2})^{2}}{4c_{0}} \alpha^{2}.
\end{equation}
Substituting the factor $c_{0}$ \eqref{eq:c0}, the eigenvalue collision point
is estimated as
\begin{equation}
  \kappa_{\alpha}^{\rm col}
  = - \dfrac{K_{0}^{\rm c}\pi^{4}}{2\beta_{4}^{2}} \alpha^{2}.
\end{equation}
The values $\beta_{4}=3$ and $K_{0}^{\rm c}\simeq 0.986225$ give 
\begin{equation}
  \kappa_{\alpha}^{\rm col} \simeq -5.34 \alpha^{2}.
\end{equation}
The jump point $\kappa_{\alpha}^{\rm J}$ is
\begin{equation}
  \kappa_{\alpha}^{\rm J}
  \simeq \dfrac{2(\wt{L}_{3/2})^{3/2}}{3[3\wt{L}_{5/2}F_{0}^{(4)}(0)]^{1/2}} |\alpha|^{3/2}
\end{equation}
at leading order. We have
\begin{equation}
  F_{0}^{(4)}(0) = - 6 A \beta_{4}^{2}
  = - \dfrac{6\beta_{4}^{5/2}}{\sqrt{2}\pi\Gamma(1/4)}
  \simeq -5.80642.
\end{equation}
Therefore, using \eqref{eq:wtL3252}, we have
\begin{equation}
  \kappa_{\alpha}^{\rm J} \simeq 6.29 |\alpha|^{3/2}.
\end{equation}


\begin{thebibliography}{99}

\bibitem{Morrison}
  P. J. Morrison,
  Hamiltonian and action principle formulations of plasma physics,
  Phys. Plasmas {\bf 12}, 058102 (2005).


\bibitem{Vlasov}
  A. A. Vlasov,
  On high-frequency properties of electron gas,
  Journal of Experimental and Theoretical Physics, {\bf 8}, 291 (1938).

\bibitem{Landau} L. D. Landau, On the vibrations of the electronic plasma, J. Phys. USSR {\bf 10}, 25 (1946).
  

\bibitem{ONeil-Winfrey-Malmberg-71}
  T. M. O'Neil,  J. H. Winfrey, and J. H. Malmberg,
  Nonlinear interaction of a small cold beam and a plasma,
  Phys. Fluids {\bf 14}, 1204 (1971).

\bibitem{delCastilloNegrete-98}
  D. del-Castillo-Negrete,
  Nonlinear evolution of perturbation in marginally stable plasmas,
  Phys. Lett. A {\bf 241}, 99 (1998).


\bibitem{ElskensBook}
  Y. Elskens and D. Escande,
  {\it Microscopic dynamics of plasmas and chaos}
  (Institute of Physics Publishing, Bristol, 2003). 
  
\bibitem{Balmforth-Morrison-Thiffeault-13}
  N. J. Balmforth, P. J. Morrison, and J.-L. Thiffeault,
  Pattern formation in Hamiltonian systems with continuous spectra; A normal-form single-wave model,
  arXiv:1303.0065.
  
\bibitem{Crawford-94}
  J. D. Crawford,
  Universal trapping scaling on the unstable manifold for a collisionless electrostatic mode,
  Phys. Rev. Lett. {\bf 73}, 656 (1994).

\bibitem{Crawford-95}
  J. D. Crawford,
  Amplitude equations for electrostatic waves: Universal singular behavior in the limit of weak instability,
  Phys. Plasmas {\bf 2}, 97 (1995).

\bibitem{Balmforth-12}
  N. J. Balmforth, A. Roy, and C. P. Caulfield,
  Dynamics of vorticity defects in stratified shear flow,
  J. Fluid Mech. {\bf 694}, 292 (2012).

\bibitem{Wiggins}
  S. Wiggins,
  {\it Introduction to Applied Nonlinear Dynamical Systems and Chaos, 2nd ed.}
  (Springer-Verlag, New York, 2003).

\bibitem{Antoniazzi-etal-07}
  A. Antoniazzi, D. Fanelli, S. Ruffo, and Y. Y. Yamaguchi,
  Nonequilibrium tricritical point in a system with long-range interactions,
  Phys. Rev. Lett. {\bf 99}, 040601 (2007).
  
  \bibitem{Palmer}
  P. L. Palmer, J. Papaloizou, and A. J. Allen,
  Neighbouring Equilibira to Radially Anisotropic Spheres-Possible End-States for Violently Relaxed Stellar Systems,
  Mon. Not. R. Astron. Soc. {\bf 246}, 415 (1990).

\bibitem{Barre-Metivier-Yamaguchi-16} J. Barr\'e, D. M\'etivier, Y.Y.  Yamaguchi,  Trapping scaling for bifurcations in the Vlasov systems. Physical Review E, 93(4), 042207 (2016).
  
 \bibitem{Barre-Metivier-Yamaguchi-20}
  J. Barr{\'e}, D. M{\'e}tivier, and Y. Y. Yamaguchi,
  Towards a classification of bifurcations in Vlasov equations,
  Phys. Rev. E {\bf 102}, 052208 (2020).
  
\bibitem{Porras-Cirac-04}
  D. Porras and J. I. Cirac,
  Effective quantum spin systems with trapped ions,
  Phys. Rev. Lett. {\bf 92}, 207901 (2004).
  
\bibitem{Kim-etal-09}
  K. Kim, M.-S. Chang, R. Islam, S. Korenblit, L.-M. Duan, and C. Monroe,
  Entanglement and tunable spin-spin couplings between trapped ions using multiple transverse modes,
  Phys. Rev. Lett. {\bf 103}, 120502 (2009).

\bibitem{Britton-etal-12}
  J. W. Britton, B. C. Sawyer, A. C. Keith, C.-C. J. Wang, J. K. Freericks, H. Uys, M. J. Biercuk, J. J. John, and J. Bollinger,
  Engineered two-dimensional Ising interactions in a trapped-ion quantum simulator with hundreds of spins,
  Nature (London) {\bf 484}, 489 (2012).

\bibitem{Islam-etal-13}
  R. Islam, C. Senko, W. C. Campbell, S. Korenblit, J. Smith, A. Lee, E. E. Edwards, C.-C. J. Wang, J. K. Freericks, and C. Monroe,
  Emergence and frustration of magnetism with variable-range interactions in a quantum simulator,
  Science {\bf 340}, 583 (2013).

\bibitem{Richerme-etal-14}
  P. Richerme, Z.-X. Gong, A. Lee, C. Senko, J. Smith, M. Foss-Feig, S. Michalakis, A. V. Gorshkov, and C. Monroe,
  Non-local propagation of correlations in quantum systems with long-range interactions,
  Nature (London) {\bf 511}, 198 (2014).

\bibitem{Leoncini-VanDenBerg-Fanelli-09}
  X. Leoncini, T. L. Van Den Berg, and D. Fanelli,
  Out-of-equilibrium solutions in the XY-Hamiltonian mean-field model,
  EPL {\bf 86}, 20002 (2009).

\bibitem{deBuyl-Mukamel-Ruffo-11}
  P. de Buyl, D. Mukamel, and S. Ruffo,
  Self-consistent inhomogeneous steady states in Hamiltonian mean-field dynamics,
  Phys. Rev. E {\bf 84}, 061151 (2011).
  
\bibitem{Ogawa-Yamaguchi-14}
  S. Ogawa and Y. Y. Yamaguchi,
  Nonlinear response for external field and perturbation in the Vlasov system,
  Phys. Rev. E {\bf 89}, 052114 (2014).

\bibitem{Ogawa-Yamaguchi-15}
  S. Ogawa and Y. Y. Yamaguchi,
  Landau-like theory for universality of critical exponenents in quasistationary states of isolated mean-field systems.
  Phys. Rev. E {\bf 91}, 062108 (2015).

\bibitem{Tacu-Benisti-22}
  M. Tacu and D. B{\'e}nisti,
  Nonlinear adiabatic electron plasma waves: I. General theory and nonlinear frequency shift,
  Phys. Plasmas {\bf 29}, 052108 (2022).
  

\bibitem{Inagaki-Konishi-93}
  S. Inagaki and T. Konishi,
  Dynamical stability of a simple model similar to self-gravitating systems,
  Publ. Astron. Soc. Jpn. {\bf 4}, 733 (1993).

\bibitem{Antoni-Ruffo-95}
  M. Antoni and S. Ruffo,
  Clustering and relaxation in Hamiltonian long-range dynamics,
  Phys. Rev. E {\bf 52}, 2361 (1995).

\bibitem{Braun-Hepp-77}
  W. Braun and K. Hepp,
  The Vlasov dynamics and its fluctuations in the $1/N$ limit of interacting classical particles,
  Commun. Math. Phys. {\bf 56}, 101 (1977).
  
\bibitem{Dobrushin-79}
  R. L. Dobrushin,
  Vlasov equations,
  Funct. Anal. Appl. {\bf 13}, 115 (1979).

\bibitem{Spohn-91}
  H. Spohn,
  {\it Large Scale Dynamics of Interacting Particles}
  (Springer-Verlag, Heidelberg, 1991).
  
\bibitem{Barre-Olivetti-Yamaguchi-10}
  J. Barr{\'e}, A. Olivetti, and Y. Y. Yamaguchi,
  Dynamics of perturbations around inhomogeneous backgrounds in the HMF model,
  J. Stat. Mech. (2010) P08002.
  
\bibitem{Yamaguchi-Ogawa-15}
  Y. Y. Yamaguchi and S. Ogawa,
  Conditions for predicting quasistationary states by rearrangement formula,
  Phys. Rev. E {\bf 92}, 042131 (2015).

\bibitem{Barre-Yamaguchi-09}
  J. Barr{\'e} and Y. Y. Yamaguchi,
  Small traveling clusters in attractive and repulsive Hamiltonian mean-field models,
  Phys. Rev. E {\bf 79}, 036208 (2009).

\bibitem{deBuyl-10}
  P. de Buyl,
  Numerical resolution of the Vlasov equation for the Hamiltonian Mean-Field model,
  Commun. Nonlinear Sci. Numer. Simulat. {\bf 15}, 2133 (2010).
  


\end{thebibliography}
\end{document}